\documentclass[twocolumn,superscriptaddress,prd,showkeys,showpacs,nofootinbib]{revtex4-1}

\usepackage{graphicx}
\usepackage[colorlinks = true,
            linkcolor = cyan,
            urlcolor  = blue,
            citecolor = red,
            anchorcolor = blue]{hyperref}\usepackage{color}
\usepackage{amssymb}
\usepackage[nointegrals]{wasysym}
\usepackage{amsthm}
\usepackage{textcomp}
\usepackage{mathtools}

\usepackage[normalem]{ulem}

\usepackage{lineno}

\usepackage{comment}
\usepackage[separate-uncertainty,retain-explicit-plus,per-mode = symbol]{siunitx}
\usepackage{url}

\newcommand{\iso}[2]{\mbox{$^{#1}$#2}}

\begin{document}

\title{Search for the Migdal effect in liquid xenon with keV-level nuclear recoils}

\author{J.~Xu} \email[Corresponding author, ] {xu12@llnl.gov}\affiliation{Lawrence Livermore National Laboratory, Livermore, CA 94551, USA}  
\author{D.~Adams} \affiliation{C.N. Yang Institute for Theoretical Physics, Stony Brook University, Stony Brook, NY 11794, USA}  
\author{B.~Lenardo} \email[Corresponding author, ] {blenardo@slac.stanford.edu} \affiliation{SLAC National Accelerator Laboratory, Menlo Park, CA 94025, USA}  
\author{T.~Pershing} \affiliation{Lawrence Livermore National Laboratory, Livermore, CA 94551, USA}  
\author{R.L.~Mannino} \affiliation{Lawrence Livermore National Laboratory, Livermore, CA 94551, USA}  
\author{E.~Bernard} \affiliation{Lawrence Livermore National Laboratory, Livermore, CA 94551, USA}  
\author{J.~Kingston} \affiliation{Department of Physics, University of California Davis, Davis, CA 95616, USA}  \affiliation{Lawrence Livermore National Laboratory, Livermore, CA 94551, USA}
\author{E.~Mizrachi} \affiliation{Physics Department, University of Maryland, College Park MD 20742, USA}  \affiliation{Lawrence Livermore National Laboratory, Livermore, CA 94551, USA}
\author{J.~Lin} \affiliation{Department of Physics, University of California Berkeley, Berkeley, California 94720, USA}  
\author{R.~Essig} \affiliation{C.N. Yang Institute for Theoretical Physics, Stony Brook University, Stony Brook, NY 11794, USA} 
\author{V.~Mozin} \affiliation{Lawrence Livermore National Laboratory, Livermore, CA 94551, USA}  
\author{P.~Kerr} \affiliation{Lawrence Livermore National Laboratory, Livermore, CA 94551, USA}  
\author{A.~Bernstein} \affiliation{Lawrence Livermore National Laboratory, Livermore, CA 94551, USA}  
\author{M.~Tripathi} \affiliation{Department of Physics, University of California Davis, Davis, CA 95616, USA}

\date{\today}

\begin{abstract}
  The Migdal effect predicts that a nuclear recoil interaction can be accompanied by atomic ionization, allowing many dark matter direct detection experiments to gain sensitivity to sub-GeV masses. We report the first direct search for the Migdal effect for M- and L-shell electrons in liquid xenon using \SI{7.0\pm1.6}{keV} nuclear recoils produced by tagged neutron scatters. Despite an observed background rate lower than that of expected signals in the region of interest, we do not observe a signal consistent with predictions. We discuss possible explanations, including inaccurate predictions for either the Migdal rate or the signal response in liquid xenon.  We comment on the implications for direct dark-matter searches and future Migdal characterization efforts.
\end{abstract}

\maketitle




\emph{Introduction --} 
Direct detection dark matter experiments have not observed definitive evidence of dark matter despite dramatic sensitivity improvements to the keV-scale nuclear recoils (NRs) expected from generic GeV--TeV particle candidates~\cite{LZ2022_FirstResult, XenonNT2023_NRDM, PandaX2021_DM, DarkSide2018_532day, CDMS2018_DM}. 
Experiments are now expanding their searches to include possible low-energy signals from light ($<$\SI{1}{GeV/c^2})  dark matter candidates motivated by various dark-sector models~\cite{Essig2013_DarkSector,Essig2017_CVLowMassDM,Essig2022_LowMassDM}. 
While these light dark matter particles would still transfer energy to nuclei or electrons through scattering, 
the energy transfers in such processes are below the few-keV thresholds of many current experiments. 
Observation of these interactions requires detectors' energy thresholds to be lowered into the sub-keV or even eV energy region. 

One possible avenue for detecting low-mass dark matter with current technology is through the Migdal effect~\cite{Migdal_1941}. 
It is predicted that the displacement of a nucleus relative to the electron shells in a scattering process can -- with a small probability -- excite or ionize the atom. The subsequent atomic relaxation would then produce additional energy depositions. 
This effect was first discussed in the context of dark matter searches in Ref.~\cite{Moustakidis:2005gx,Vergados:2005dpd,Bernabei:2007jz} 
and then further developed~\cite{Ibe2017_Migdal, Liu2020_Migdal,Bell2020_Migdal,Essig2020:Migdal,Adams2022_Migdal,Baxter2020:Migdal,Cox2022_Migdal,Berghaus:2022pbu}. For the interactions of sub-GeV dark matter that would produce NRs only in the sub-keV energy region in typical targets, the Migdal signals 
can contain electron recoils (ERs) at the keV-scale~\cite{Ibe2017_Migdal,Essig2020:Migdal,Baxter2020:Migdal,Bell2020_Migdal} -- above the thresholds of many existing detectors. 
For some dark matter models, Migdal interactions even dominate over direct dark matter-electron scatters~\cite{Baxter2020:Migdal,Essig2020:Migdal}.  
Therefore, despite a low probability ($\mathcal{O}(10^{-5})$\ relative to NR), Migdal interactions can greatly enhance a detector's sensitivity to low-mass dark-matter.
By incorporating this effect into their signal models, several experiments have reported substantially improved -- and in some cases, world-leading -- sensitivity to sub-GeV dark matter~\cite{LUX2019:Migdal,SENSEI:2020dpa,CDMS2022_Migdal,XENON1T2019:Migdal,DarkSide2022:Migdal,Berghaus:2022pbu}. 
A range of ideas for measuring the NR-induced Migdal effect have been proposed~\cite{Nakamura2020_Migdal,UKMIGDAL2022, Bell2022_Migdal,Adams2022_Migdal}, but it remains to be observed in any detector media to date. 

In this work we report the first direct search for the co-production of Migdal ER signals accompanying nuclear recoils in liquid xenon (LXe). 
Dual-phase xenon time projection chambers (TPCs) are widely used in dark matter searches due to their ability to achieve keV-scale thresholds (and below) with large target masses and low backgrounds~\cite{XENONnT2022_ER,LZ2022_FirstResult,XENON1T2019:Migdal}. 
These detectors measure both scintillation and ionization signals (termed S1 and S2, respectively)
 from particle interactions, which are proportional to the deposited energy and whose ratio
enables discrimination between ER and NR energy depositions. 
We take advantage of the latter to search for Migdal events with sufficient ER energy to be well-separated from low-energy NR events produced by fixed-angle scatters of \SI{14.1}{MeV} neutrons in a compact xenon TPC. 

\begin{figure}[!t]
    \centering
    \includegraphics[width=0.49\textwidth]{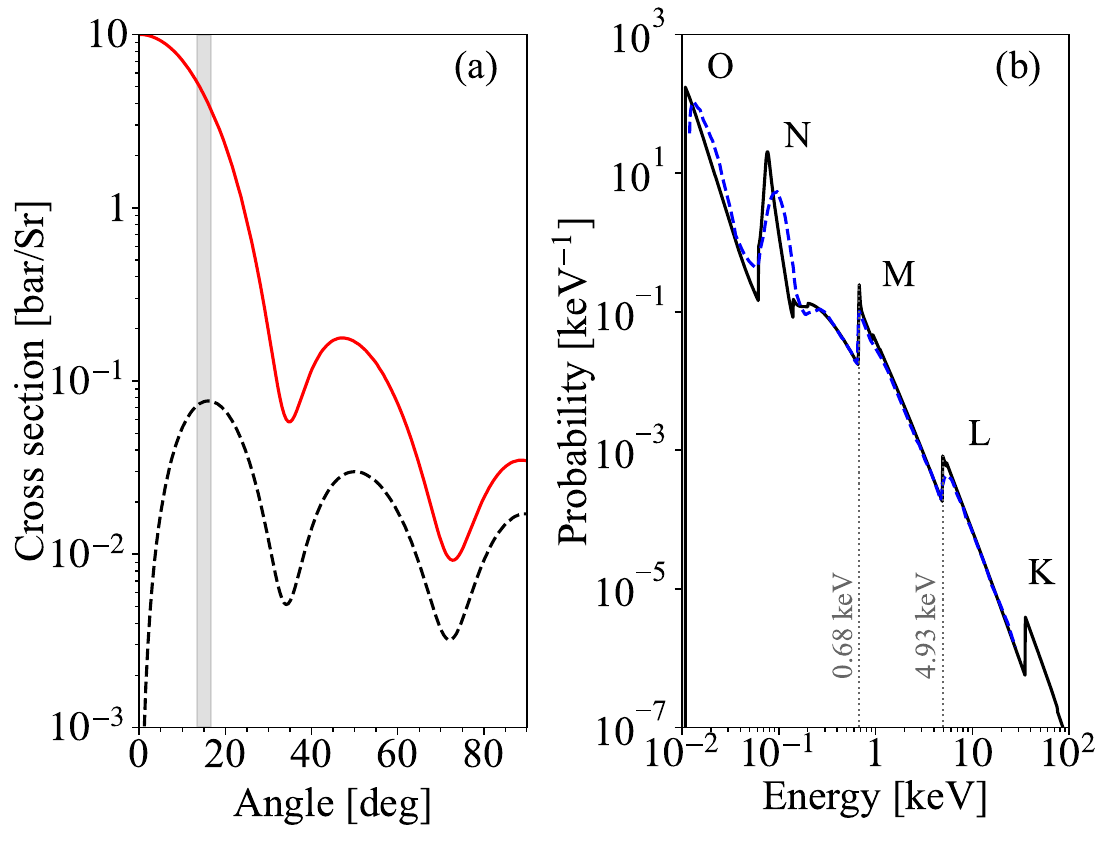}
    \caption{{\bf(a)} Calculated angular cross section for Migdal interactions (dashed, black) and elastic scatters (solid, red)  between \SI{14}{MeV} neutrons and natural xenon. The shaded region indicates the $\pm1\sigma$ range of neutron scattering angles used in this experiment. {\bf (b)} Predicted spectrum of the total ER energy released by Migdal interactions in the present work using first-principle inputs (solid, black) and photoabsorption data (dashed, blue). The energies of the peaks associated with the xenon M- and L-shells are labeled in grey.}
    \label{fig:cs}
\end{figure}

\emph{Theory --} 
Figure~\ref{fig:cs}(a) shows the calculated Migdal cross section, following the method in~\cite{Ibe2017_Migdal}, for interactions of \SI{14}{MeV} neutrons with natural xenon, in comparison to that of elastic n-Xe scatters. 
The Migdal cross section approximately follows  that of elastic scatters but is additionally modulated by a term proportional to $E_n(1-\cos\theta)$, where $E_n$ is the incoming neutron energy and $\theta$ is the scattering angle~\cite{Adams2022_Migdal}. 
Higher neutron energies increase the Migdal interaction rate and reduce backgrounds from neutron multiple-scatters (NMS).
The Migdal cross section peaks at a neutron scattering angle of $\theta \sim$\ang{16}, where its ratio to the elastic cross section is $\sim$1/60; while this ratio gradually increases with the scattering angle, the lower absolute signal rate and the higher energy of the background NRs make Migdal searches difficult at larger angles. 
Normalizing the Migdal interaction strength to that of elastic scatters reduces the dependence of the predicted signal rate on the n-Xe scattering cross section and detector efficiencies.  

Figure~\ref{fig:cs}(b) (solid line) shows the predicted distribution of total Migdal ER energy, which includes the energy deposited by the initial Migdal electron and the subsequent atomic relaxations. 
The spectral shape is relatively insensitive to the neutron energy and scatter angle.
For signals considered in this work, the Migdal excitation probability is 3--4 orders of magnitude  below that of the Migdal ionization probability~\cite{Ibe2017_Migdal} and is thus neglected. 
The probability for a K, L, and M shell ionization
in a Migdal interaction is approximately 5$\times$10$^{-5}$, 2$\times$10$^{-3}$, and 4$\times$10$^{-2}$, respectively. 
Although the Migdal probability increases for outer (N \& O) shells, these interactions deposit too little ER energy to be identified in our experiment. 

We also carried out a second calculation using experimental photoabsorption data based on its similarity to the Migdal process, which allows an atomic data-driven prediction of the Migdal rate and spectrum that are less sensitive to uncertainties from modeling of atomic physics~\cite{Liu2020_Migdal}.
The spectrum is also shown in Figure~\ref{fig:cs}(b), which agrees with the first-principles calculation within 20\%. 
Higher-order ionization effects~\cite{Cox2022_Migdal} change the prediction by $<$5\%. 
Our analysis uses the average signal rate over these different calculations, and we assign a combined uncertainty of $\pm$11\%.

\emph{Experiment --} 
The measurement was carried out at Lawrence Livermore National Laboratory using the XeNu xenon TPC. 
The active LXe volume of \diameter \SI{3.8}{cm}$\times$\SI{2.5}{cm} was enclosed in a PTFE reflector cylinder and is viewed by four Hamamatsu R8520-406 PMTs from the top and one Hamamatsu R8778 PMT from the bottom. 
An electric field cage consisting of three stainless steel grids (cathode, extraction grid and anode) 
and three voltage step-down rings between the cathode and the extraction grids provided a drift field of \SI{200\pm42}{V/cm} and a liquid extraction field of \SI{6.6}{kV/cm} (\SI{12.2}{kV/cm} in the gas). 
The high electric field leads to an electron extraction efficiency of 95$\pm$3\% from the liquid into the gas and an observed S2 ionization signal gain of 71.5$\pm$1.5~photoelectrons (PHEs) per extracted electron. 
The collection efficiency for the S1 scintillation light in LXe was measured to be \SI{0.127\pm 0.019}{PHE/photon}. 

The neutron source consisted of a deuterium-tritium (DT) neutron generator placed inside a shielding structure of up to \SI{25}{cm} of lead surrounded by up to \SI{1}{m} of borated water~\cite{Pershing2022_XeNR}. The neutrons were collimated by a \SI{2.5}{cm} square opening in both the lead shielding and a \SI{1}{m} long borated polyethylene (30\%-by-weight) structure fitted into the water shielding. The transmitted neutrons were emitted at a \ang{90} angle relative to the axis of the generator, where it produces monochromatic \SI{14.1}{MeV} neutrons with little sensitivity to its operating conditions. 

The xenon TPC was centered on the beam at a distance of \SI{1.5}{m} from the source. Up to \SI{12.7}{cm} of lead was placed between the shielding and the TPC (\SI{3}{mm} lead used at the beam opening) to attenuate secondary $\gamma$-rays from neutron interactions in the shielding. A circular array of 14 liquid scintillator (LS) detectors (EJ301/309, \diameter \SI{10}{cm}$\times$\SI{7.6}{cm}), centered on the beam axis, was placed \SI{95}{cm} behind the TPC and used to tag neutrons scattered at \ang{15.0}, with a $1\sigma$ spread of $\pm$\ang{1.6}. 
In this configuration, the NR energy deposited by elastic neutron scatters is \SI{7.0\pm1.6}{keV}, which on average produces $\sim$50 scintillation photons and $\sim$50 electrons.
For every 100,000 NRs of this energy, around 1350 Migdal interactions may be produced, of which 55 (2.3) are from the M (L) shell. 
An M(L)-shell xenon Migdal signal alone 
can produce up to 20 (200) photons and $\sim$50 (175) electrons in LXe, allowing them to be well-separated from pure NR events in the S1-S2 signal space. 

Data acquisition was triggered by a signal in any neutron detector followed by an S2 signal in the TPC within \SI{30}{\us}. 
The TPC threshold was set such that trigger efficiency approached 100\% for S2$>$\SI{12}{e^-}. The thresholds of the neutron detectors were set at $\sim$2\% of the maximum observed proton recoil energy. 
We digitized every PMT output at \SI{250}{MS/s} for \SI{30}{\us} before and after the trigger, 
providing an event window substantially longer than the maximum electron drift time in the TPC (\SI{17.7}{\us}). 
With an average neutron rate of \SI{2.7e7}{n/s} and a total operation time of $\sim$100 hours, we have approximately 2.2$\times10^8$ neutrons incident on the active xenon volume with around 600,000 neutron coincidences recorded.  

\emph{Analysis --} 
The Migdal signal searches are based on the selection of fixed-angle, single-scatter (SS) neutron interactions in the TPC. 
We utilize the powerful neutron/$\gamma$-ray pulse shape discrimination (PSD) of the LS detectors to identify neutron candidates with a PSD parameter 
 $>5\sigma$ away from the $\gamma$-ray PSD distribution. 
For each neutron coincidence event, pulses in the TPC are classified as S1s or S2s using their pulse widths, which are $\mathcal{O}$(\SI{10}{ns}) for S1s and $\mathcal{O}$(\SI{1}{\us}) for S2s. 
For candidate S2 pulses, we apply additional quality cuts on the ratio of signal strength in the top/bottom PMTs, the asymmetry of the pulses' rise and fall times, and the goodness-of-fit when the pulses are fitted to a Gaussian; each cut is set at $>$98\% acceptance. 
For S1s, we use two different algorithms to reconstruct each pulse. 
For S1s of $>$\SI{15}{PHE}, such as those expected for L-shell Migdal signals, a pulse shape-based
algorithm works adequately.
However, the majority of events in this work have S1s of $<$\SI{10}{PHE}  composed of individual PHE peaks, and the efficiency of the pulse-shape-based algorithm drops. 
We therefore also use a time-based algorithm in which the S1 signal is reconstructed as all the TPC PHE signals within the $[\SI{-100}{ns},\SI{+200}{ns}]$ window around the neutron detection time in the LS. 
Events with S2-like pulses or excess PHE noise near the S1 search window are excluded to avoid S1 misidentification. 
If a candidate S1 contains more than one peak, we apply a quality cut on the maximum time between the first two peaks to suppress random PHE pile-ups. 
The cut value varies with S1~size to maintain a constant 
efficiency for S1s in the range 4--\SI{40}{PHE}. 

With the LS and TPC pulses identified, another cut is applied on the scattered neutron time of flight (TOF), defined as the difference between the TPC S1 and the LS neutron time, to remove accidental coincidences and ``passive scattering'' (PS) -- events in which the neutron deposits energy in other parts of the experimental setup than the TPC and the LS detector.
Events outside the TOF window also provide an estimate of backgrounds from misidentified S1s and accidental coincidences between TPC events and unrelated LS neutron interactions. 

Finally, we require that the event contain a single S1-S2 pair to remove NMS events in the TPC. 
In addition, we reject the widest 10\% of S2 pulses (as a function of the drift time), which targets NMS events in which the multiple vertices are too close to resolve.

\begin{figure}[!t]
    \centering
    \includegraphics[width=0.49\textwidth]{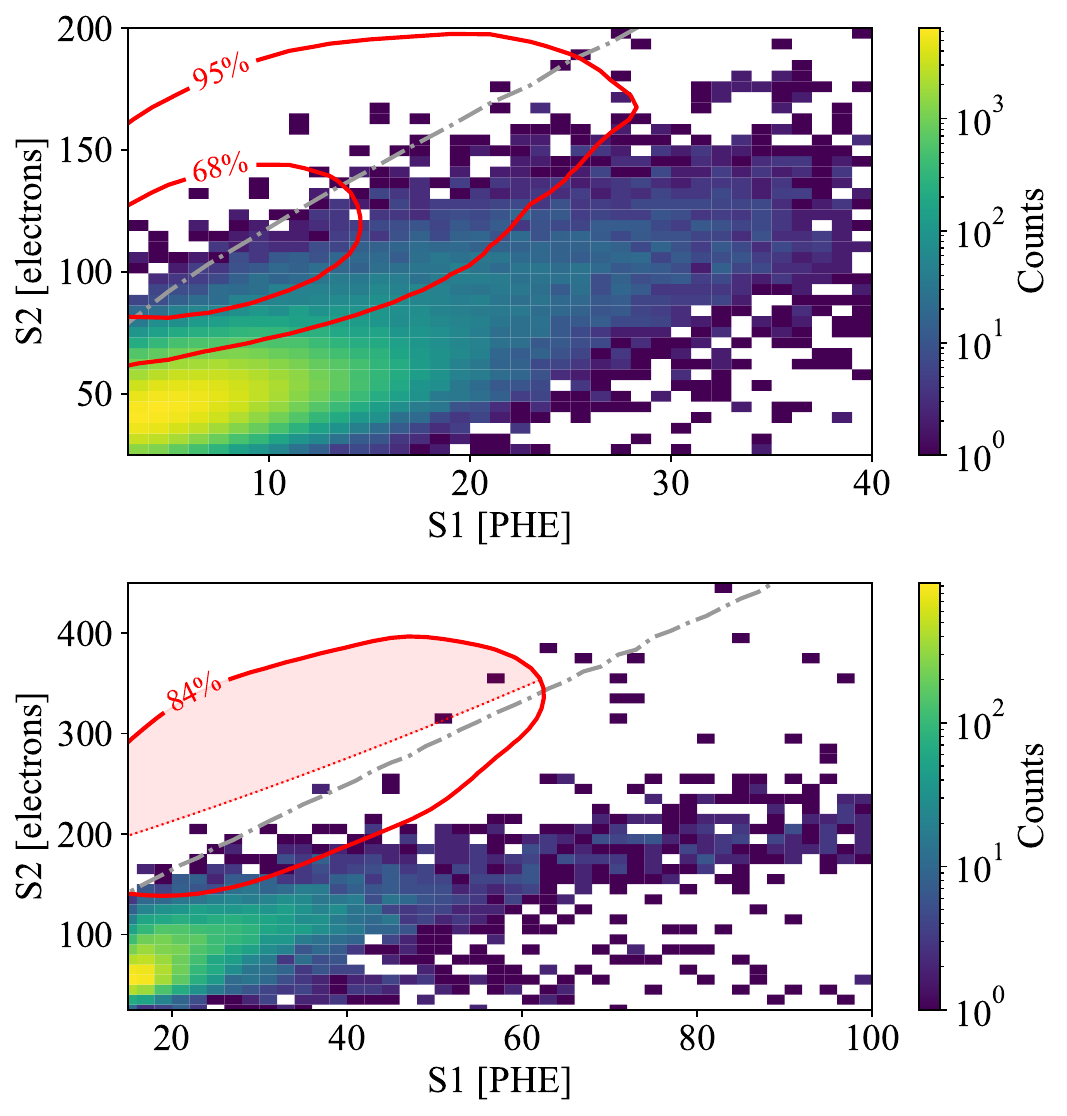}
    \caption{The S2 vs S1 distributions for the data used in the M-shell (top) and L-shell (bottom) Migdal signal searches. 
    In both figures, signal contours (solid red) and the ER distribution median (dot-dashed grey) are also plotted. 
    The shaded region in the bottom illustrates the analysis ROI for the L-shell Migdal search. 
    }
    \label{fig:s12}
\end{figure}

Our analysis primarily focuses on the M-shell signals, for which we consider events with 4$<$S1$<$\SI{40}{PHE} and 25$<$S2$<$\SI{200}{e}$^-$. This analysis employs the time-based S1 reconstruction algorithm for all S1s and uses a TOF cut window of \SI{55}{ns}.
Figure~\ref{fig:s12} (top) shows the S1-S2 distribution for 
the $\sim$300,000 TPC events passing all cuts, the majority of which are SS NRs. 
Signals and backgrounds are modeled by simulating the entire experimental setup with Geant4~\cite{Geant4_2016}. The resulting energy depositions are processed with the NEST software package~\cite{szydagis_m_2022_6448408}, which models scintillation and ionization yields to predict the S1-S2 distributions. The simulations provide three separate NR background distributions: neutron SS, PS, and NMS.
We vary a selected set of NEST parameters to match the NR models to the high-statistics SS NR peak and the higher-energy PS/NMS events where Migdal contributions are predicted to be negligible. 
The NMS model is validated against resolved NMS events, which have two S2s separated by over 1mm vertically, and yields a good agreement. 
In addition, we include an ER background model expected of Compton scatters of MeV-scale $\gamma$-rays from neutron-induced reactions. This model is constructed by fitting NEST to Compton scatter events induced by a \iso{60}{Co} $\gamma$-ray calibration source placed next to the TPC. 
Following the practices in \cite{Ibe2017_Migdal} and \cite{Bell2020_Migdal}, we treat the LXe responses to the NR and ER components of a Migdal interaction as independent; the signal model is created by adding the S1 and S2 signals from an ER event sampled from the distribution in Figure~\ref{fig:cs}(b) to those of the simulated NR, scaled by the ratio given in Figure~\ref{fig:cs}(a). 
The 1$\sigma$ and 2$\sigma$ contours for Migdal events with ER energies between 0.5~--~\SI{3}{keV} are plotted over data in Figure~\ref{fig:s12} (top, solid lines, red). 
Thanks to the low energy of SS NRs and the large number of additional S2 electrons expected of Migdal ERs, the predicted signal region is significantly separated from the pure NR population. 

A two-dimensional maximum likelihood fit was carried out in two steps: first, the entire region of 4$<$S1$<$\SI{40}{PHE} and 25$<$S2$<$\SI{200}{e^-} was fit with the Migdal, SS, PS, NMS, and ER background rates all allowed to float; then, the SS rate was fixed, and a second fit was performed with the S2 lower bound increased from \SI{25}{e^-} to \SI{80}{e^-} and all other components floating. 
By excluding the data region dominated by pure NRs and low-energy Migdal interactions, the second fit 
is expected to be less susceptible to imperfect modeling of the NR distribution. 

\begin{figure}[!t]
    \centering
    \includegraphics[width=0.49\textwidth]{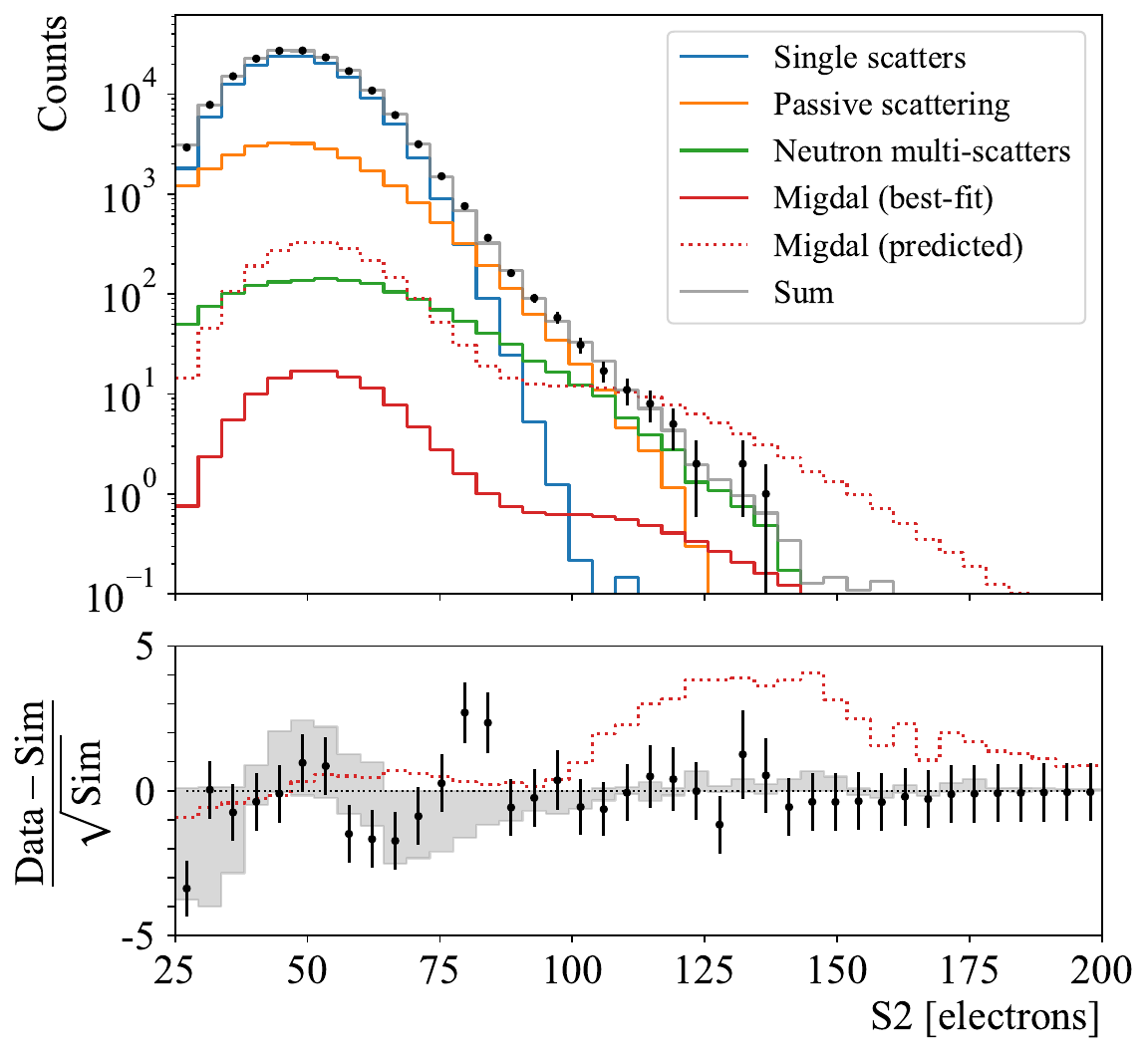}
    \caption{Projected S2 spectrum for events with 4$<$S1$<$\SI{10}{PHE} (black dots), along with the best-fit simulated distributions for SS (blue), PS (orange) and NMS (green) backgrounds and the Migdal signal (red). The summed best fit model is shown in grey. We also show the Migdal signal at the predicted rate (dashed red) for comparison.
    Residuals are shown for both the best-fit model and for the the model with the Migdal rate fixed at its predicted value. The shaded band represents the systematic uncertainty in the NR background models.
    }
    \label{fig:mfit}
\end{figure}

The best-fit Migdal signal rate is found to be 16.3$^{+21.7}_{-16.3}\,$ counts, where the range represents the statistical 68\% confidence interval computed using the profile likelihood ratio technique. This is an order of magnitude lower than the predicted value of 148.2$\pm$16.3
and is consistent with 0. 
This tension is illustrated visually in Figure~\ref{fig:mfit} where the 
nominal Migdal signal model (dotted, magenta) is contrasted with the best fit (solid, red) for the S1 range of 4--\SI{10}{PHE}, and the nominal model alone already over-predicts the observed event rate for a range of S2 values. 
The dominant backgrounds for S2$>$\SI{100}{e^-} are from NMS and PS, while fluctuations of SS NRs play an insignificant role. 
Further, the estimated background rates in this region are below the expected Migdal rate, suggesting an expected signal to noise ratio (SNR) of $>$1. Additional information on the model fits can be found in the Supplemental Material. 

Systematic uncertainties in the NR models are evaluated by varying the NR NEST parameters to produce models for which the fitted likelihood was up to 
2$\sigma$ worse than the global best-fit; the resulted spread is illustrated by the shaded grey band in Figure~\ref{fig:mfit} and yielded a conservative uncertainty in the fitted Migdal rate of $\pm$8.1 counts.  
Uncertainty in the Migdal ER model component is evaluated by varying the NEST ER yields within their reported uncertainty band~\cite{Szydagis:2022ikv}, producing an uncertainty of $\pm$6.3 counts.  
The combination of these uncertainties does not significantly change the analysis result, demonstrating the robustness of this work against modeling inaccuracies. 

We also extend our analysis window to attempt a L-shell Migdal signal search. 
Thanks to the large S1 sizes of expected signals, the simple pulse shape-based S1 identification algorithm is used, allowing us to relax the S1 cleanliness cuts, which boosts the NR statistics by $\sim30\%$. The more precise S1 timing also enables a more stringent TOF cut (\SI{19}{ns}), leading to stronger rejection of accidental backgrounds. The resulting dataset is shown in Figure~\ref{fig:s12} (bottom).
In this high energy region, up to 10\% of S2 saturation and up to 5\% of relative efficiency loss from analysis cuts are expected, so 
we perform a simple cut-and-count analysis. The signal region of interest (ROI) is defined, prior to quantitative analysis, using the predicted distribution for Migdal events with ER energy above \SI{3}{keV}; we accept events above the median but within the 84\% contour, illustrated by the shaded region in Figure~\ref{fig:s12} (bottom). 
We observe 2 events in this ROI, while expecting 5.6$\pm$1.2 signals. Backgrounds are estimated using the simulated event distributions described above: the Compton ER background model, scaled to match the observed rates in a control region of 100$<$S1$<$\SI{150}{PHE} (above ER median), yields an estimate of 2.1$\pm$0.9 counts in the ROI; the PS and NMS background rates are evaluated to be $<$0.1 count. 
A tension between the prediction and observation is observed with a nominal SNR of $>$1, but due to low statistics and additional sources of uncertainties, this analysis is not as powerful as the M-shell search. 

\emph{Discussion --}
The non-observation of the expected Migdal signal could suggest that current calculations over-estimate the rate of Migdal ionization in liquid xenon, at least for M-shell and L-shell electrons. If true, this could imply that dark matter searches using the Migdal effect may not achieve the sensitivities previously expected.
We note, however, that this possibility challenges theoretical expectations, which are bolstered by the agreement between the atomic data-driven and first-principles calculations and by our experimental design further mitigating nuclear physics uncertainties. 

We also consider the possibility that our null result is due to enhanced electron-ion recombination in the liquid xenon because of the close proximity of the ER and NR tracks in a Migdal event.
This would convert ionization signals into hard-to-detect scintillation signals, shifting Migdal events into the region of the S1-S2 space where our experiment has reduced sensitivities. 
This complication has not been considered in past Migdal studies. While similar phenomena have been observed in other contexts~\cite{Temples2021_127Xe, Baudis:2013cca}, 
it has not been studied in the regime relevant for this experiment,  
so we only make some qualitative statements here. 
At the energies relevant to this measurement, the $\mathcal{O}(10-\SI{100}{nm})$ track lengths of the NR and ER\footnote{The fluorescence yields of L- and M-shell transitions in Xe are $\mathcal{O}(10^{-1})$ and $\mathcal{O}(10^{-3})$ respectively~\cite{Hribar1977,McGuire1972}, meaning the ER energy from Migdal events in a liquid xenon detector is primarily deposited via low-energy electron emission.} components are smaller than the $\sim$\SI{5}{\mu m} electron thermalization radius in LXe~\cite{Mozumder1995}. We then assume that the recombination process for Migdal events is similar to that of low energy ERs, for which the recombination probability scales approximately linearly with the initial number of electron-ion pairs in the keV energy region~\cite{LUX2016_3H}. 
As the NR and ER components of a 
Migdal interaction produce comparable numbers of electron-ion pairs in our experiment, the combined signal could have a recombination probability up to 2 times larger. This could alleviate the tension between our observation and prediction. Additional experiments or a first-principles simulation of recombination in liquid xenon, similar to work done for liquid argon~\cite{Foxe2015_LowESim}, could 
yield further insight into this possibility.  

Our experiment is the first to demonstrate sufficiently low backgrounds to study the NR-induced Migdal effect in any detector medium, and provides a path forward for future characterization efforts. If enhanced recombination, instead of a reduced Migdal interaction strength, is the cause for the null result reported here, low-mass dark matter searches would still benefit from the Migdal effect because the NR component in such dark matter interactions will be negligible.
For the same reason, future LXe Migdal experiments using lower-energy NRs could reduce ambiguity and characterize this effect more conclusively.

\begin{acknowledgments}

 This project is supported by the U.S. Department of Energy (DOE) Office of Science, 
 Office of High Energy Physics under Work Proposal Number SCW1676 awarded to Lawrence Livermore National Laboratory (LLNL). 
 J.~K. was partially supported by the DOE/NNSA under Award Number DE-NA0000979 through the Nuclear Science and Security Consortium. 
R.~E. acknowledges support from DoE Grant DE-SC0009854, Simons Investigator in Physics Award 623940, and the US-Israel Binational Science Foundation Grant No.~2016153. D.~A. was supported by DoE Grant DE-SC0009854 and Simons Investigator in Physics Award 623940. 
B.~L. was supported by the DOE Laboratory Directed Research and Development program at SLAC National Accelerator Laboratory, under contract DE-AC02-76F00515 as part of the Panofsky Fellowship.

We thank Henrique Ara\'{u}jo, Eric Dahl, Rick Gaitskell, Dan McKinsey, Matthew Szydagis, Greg Blockinger and other LUX-ZEPLIN members for constructive discussions on the Migdal search. 
 We thank Chris McCabe and Peter Cox for helpful discussions with checking the uncertainties on the Migdal rate. 

This work was performed under the auspices of the U.S. Department of Energy by Lawrence Livermore
National Laboratory under Contract DE-AC52-07NA27344. 
LLNL release number: LLNL-JRNL-850170. 
\end{acknowledgments}

\bibliographystyle{apsrev}
\bibliography{biblio}

\pagebreak

\appendix

\section{Background models}
\label{sec:append_bg}
\begin{figure*}[!h]
    \centering
    \includegraphics[width=0.97\textwidth]{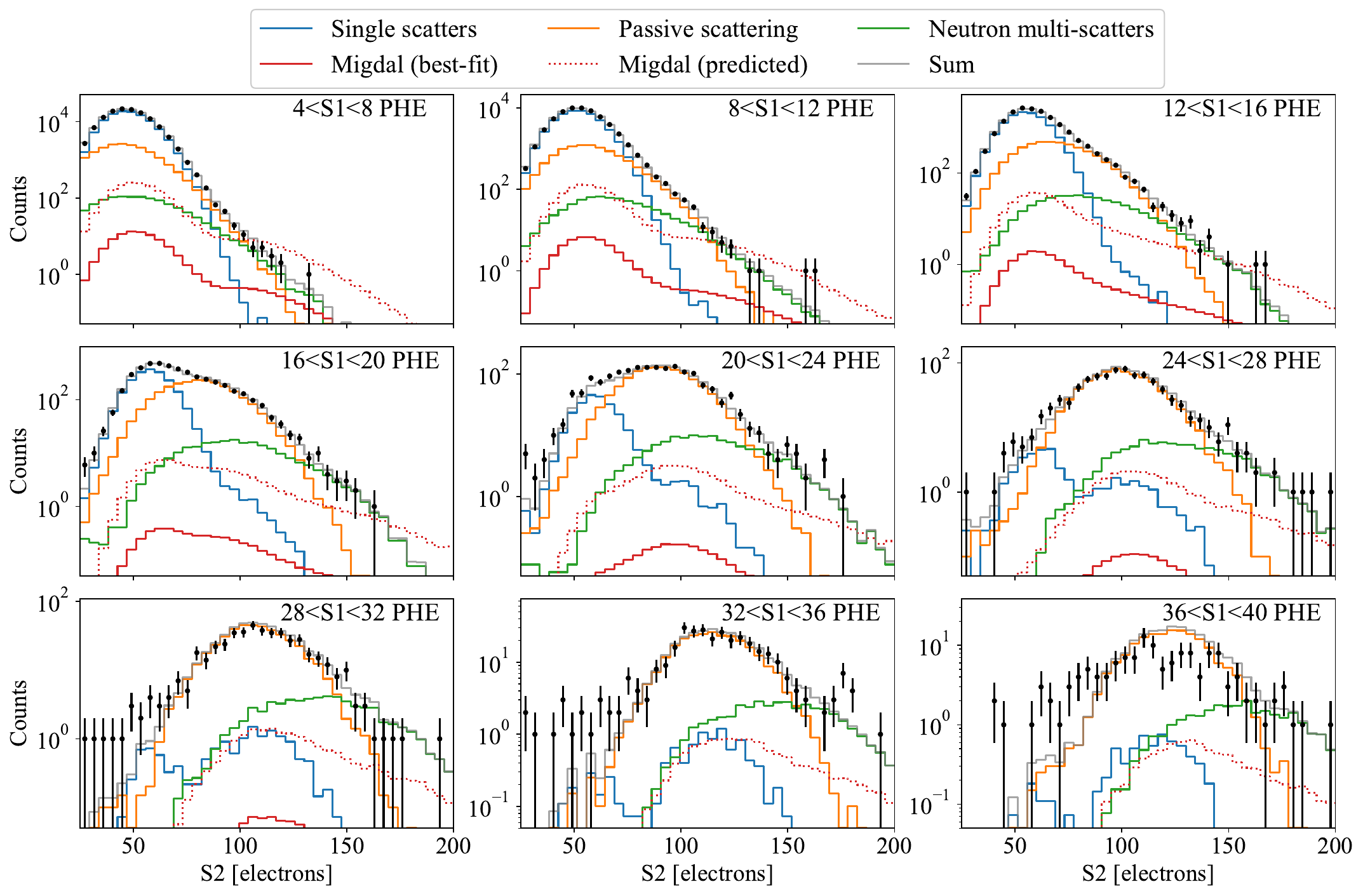}
    \caption{Measured S2 spectra in a selection of S1 slices, along with the best fit background and signal models: SS NRs (blue), TPC NMS (green), passive NMS (orange), Migdal signals (red), and summed model (gray). }
    \label{fig:bgmodel}
\end{figure*}

The most relevant background in this Migdal signal search is from the multi-scatters of neutrons in the TPC and in passive materials around the detector. 
Due to the nonlinear response of liquid xenon to NRs, multiscatter backgrounds inside the TPC can produce larger S2s than single scatters events and mimic Migdal signals. 
Multiscatter events with one TPC interaction and additional neutron scatters in the passive materials contribute broad-energy backgrounds, the high energy component of which could also pose a background to Migdal searches despite of being single-scatter NRs in the TPC. 
Figure~\ref{fig:bgmodel} shows the measured S2 energy spectra for a selection of S1 slices with \SI{2}{PHE} width, along with the best fit signal and background models. 
For S1$>$\SI{20}{PHE}, he contribution from SS NRs becomes subdominant and that from Migdal interactions is negligible; at the same time, passive and TPC neutron multiscatters dominate the observed event rate, which strongly constrains the amplitudes of these backgrounds. 
The agreement between the data and models also validates the shapes and S1 dependence of these background models. 

\onecolumngrid
\twocolumngrid

\end{document}